\documentclass[%
 twocolumn,
 amsmath,amssymb,
 aps,
 prr
 longbibliography,
 footinbib,
]{revtex4-2}

\usepackage{physics}
\usepackage{ascmac}
\usepackage{graphicx}  % Include figure files
\usepackage{dcolumn}   % Align table columns on decimal point
\usepackage{bm}        % bold math
\usepackage{physics}
\usepackage{here}
\usepackage{wrapfig}
\usepackage{color}
\usepackage[dvipsnames]{xcolor}
\usepackage{CJKutf8}
\usepackage{soul}
\usepackage{ulem}

\usepackage{hyperref}
\usepackage{xurl}
\hypersetup{breaklinks=true}

\hypersetup{
    colorlinks,
    linkcolor={red!50!black},
    citecolor={blue!50!black},
    urlcolor={blue!80!black}
}

\begin{document}
\begin{CJK}{UTF8}{min}
%\preprint{APS/123-QED}

\title{Bose--Fermi $N$-polaron state emergence from correlation-mediated blocking of phase separation}

\author{Felipe G\'omez-Lozada}
\email{feg46@pitt.edu}
\affiliation{Quantum Systems Unit, Okinawa Institute of Science and Technology Graduate University, Onna, Okinawa 904-0495, Japan}

\author{Hoshu Hiyane}
\email{hoshu.hiyane@oist.jp}
\affiliation{Quantum Systems Unit, Okinawa Institute of Science and Technology Graduate University, Onna, Okinawa 904-0495, Japan}

\author{Thomas Busch}
%\email{thomas.busch@oist.jp}
\affiliation{Quantum Systems Unit, Okinawa Institute of Science and Technology Graduate University, Onna, Okinawa 904-0495, Japan}

\author{Thom\'as Fogarty}
\email{thomas.fogarty@oist.jp}
\affiliation{Quantum Systems Unit, Okinawa Institute of Science and Technology Graduate University, Onna, Okinawa 904-0495, Japan}

\begin{abstract}

We study $N$ fermionic impurities in a one-dimensional lattice bosonic bath at unit filling. Using DMRG and mixed boundary conditions—periodic for bosons, open for fermions— we find an
$N$-polaron ground state replacing phase separation at high interspecies repulsion.
This tightly bound state of clustered particles emerges due to strong impurity-bath correlations, which induce large impurity-impurity correlations that we quantify via the von Neumann entropy and bipartite mutual information, respectively.
This system also reveals a fermionic self-localization effect from a Mott insulator background due to local correlations between the impurities and the bath. The growth of long-range correlations breaks this Mott phase, resulting in the transition to impurity clusters delocalized along the system. Finally, we show that there is a critical impurity number, which depends on intraspecies bosonic interaction, beyond which phase separation is recovered. 
\end{abstract} 

\maketitle
\end{CJK}

\section{Introduction}

Since the seminal work by Landau, impurities inside their surrounding many-body media can be understood as entities wearing a ``dress" of background excitation, leading to a notion of quasi-particles known as {\it polarons}~\cite{Landau_PZS1933, Pekar_JPU1946, Landau_JETP1948}. Studying the properties of polarons is often the first step to tackling more complicated many-body systems~\cite{Mahan_2000, Dagotto_Sci2005}, providing further insight that could lead to theoretical developments in condensed matter systems, such as the mechanism of high-temperature superconductivity~\cite{Alexandrov_SPU1992, Mott_JPCM1993, Alexandrov_2007, Muller_JSNM2014, Muller_JSNM2017, Chatterjee_2017}, one of the ultimate goals of modern many-body physics. 
As exemplified by this system, the presence of strong correlations among particles may lead to various phenomena that are fundamentally and practically important. 

The realization of polarons in cold atoms~\cite{Schirotzek_PRL2009, Kohstall_Nat2012, Koschorreck_Nat2012, Jorgensen_PRL2016, Hu_PRL2016, Grusdt_2024}, opens a new and ideal platform to study correlation effects due to the high degree of tunability and clean environment~\cite{Bloch_RMP2008, Bloch_NatP2012, Gross_Sci2017, Yan_Sci2020}.
This achievement has led to the study of the strong impurity-bath coupling regime, i.e. strongly correlated polarons, previously unattainable or uncontrollable in solid-state systems~\cite{Cetina_Sci2016, Scazza_PRL2017, Franchini_NatRM2021}, along with the investigation of the interplay between different particle statistics such as the Bose or Fermi polaron, an impurity immersed in a bosonic or fermionic bath respectively~\cite{Astrakharchik_PRA2004, Cucchietti_PRL2006, Chevy_PRA2006, Kalas_PRA2006, Bruderer_EPL2008, Prokofev_PRB2008, Tempere_PRB2009, Bei-Bing_CPL2009, Mora_PRA2009, Combescot_EPL2010, Rath_PRA2013, Li_PRA2014, Ardila_PRA2015, Levinsen_PRL2015, Volosniev_PRA2015, Grusdt_SciR2015, Christensen_PRL2015, Mistakidis_PRL2019, Mistakidis_PRR2020, Mistakidis_NJP2020, Etrych_2024}, and multiple impurities where statistics and many-body effects play an important role~\cite{Pasek_PRB2019, Ness_PRX2020, Fritsche_PRA2021, Yordanov_JPB2023, Seetharam_PRA2024, Baroni_NatP2024, Isaule_SciPC2024, Paredes_PRA2024, Guo_PRA2024}.

In this work, we study a one-dimensional lattice system that consists of a few fermionic impurities immersed in a bosonic bath at unit-filling. This system falls under the umbrella of Bose--Fermi mixtures, for which phase separation is known to occur for strong interspecies repulsion~\cite{Linder_PRA2010, Buchler_PRA2004, Viverit_PRA2000, Marchetti_PRB2008} and which is also true for some few-impurity systems~\cite{Mistakidis_PRA2019, Pasek_PRB2019, Anh-Tai_PRR2024}.
In the latter, boundary conditions (BCs) and finite size effects significantly affect the ground state of the full many-body system which is the focus of our work. 
We elucidate this by comparing open BCs for both bosonic and fermionic systems (Open(B)-Open(F) BCs) with periodic BCs for bosons while keeping the fermionic BCs open (Periodic(B)-Open(F) BCs).
While the miscible-immiscible phase transition is observed as expected for Open(B)-Open(F) BCs, for high population imbalance the Periodic(B)-Open(F) BCs system avoids phase separation at large interspecies interactions and instead forms a $N$-polaron state bounded by large bath-impurity correlations. 
We characterize this state using density profiles and appropriate correlation functions, along with tools developed in quantum information theory such as the von Neumann entropy and mutual information, which are used extensively in the characterization of quantum phase transitions~\cite{Mund_PRB2009, Mendoza-Arenas_PRA2010, Islam_Nat2015, Cornfeld_PRA2019, Perez-Romero_EPJB2021}. 

Our work is organized as follows: First, we explain the details of the Bose-Fermi model (Sec.~\ref{sec:model}), after which we illustrate the formation of the polaron and the differences that appear from mixed boundary conditions (Sec.~\ref{sec:formation}). Then, we discuss the relationship between the formation of polarons with the superfluid-insulator transition of the Bose Hubbard model (Sec.~\ref{sec:SF_MI}). We then elucidate the polaron structure at large interactions and characterize its formation using quantum information techniques (Sec.~\ref{sec:strong_regime}). Finally, we discuss the role of population imbalance (Sec.~\ref{sec:populationimbalance}) and system size (Sec.~\ref{sec:scaling}) on the mixture, and provide concluding remarks (Sec.~\ref{sec:conclusions}).

\section{Model}\label{sec:model}

We consider a mixture of bosons and spin-polarized fermions loaded in a one-dimensional optical lattice of length $L$.
The lattice depth is large enough that the low energy properties of the system can be well-described by a Bose--Fermi Hubbard model~\cite{Albus_PRA2003}
\begin{align}
    \hat{\mathcal{H}}
    &=\hat{\mathcal{H}}_{\rm B}+\hat{\mathcal{H}}_{\rm F}+\hat{\mathcal{H}}_{\rm BF}, \label{eq:Hubbard}
    \\ 
    \hat{\mathcal{H}}_{\rm B}&=-t\sum_j \left( \hat  b^\dagger_{j}\hat b_{j+1} +\rm{h.c}\right)
    +\frac{U_{\rm BB}}{2}\sum_j \hat b^\dagger_{j}\hat b^\dagger_{j}\hat b_{j}\hat b_{j},  \label{eq:Hubbard_b}\\ 
    \hat{\mathcal{H}}_{\rm F}&=-t\sum_j \left( \hat  f^\dagger_{j}\hat f_{j+1} +\rm{h.c}\right), \label{eq:Hubbard_f}
    \\ 
    \hat{\mathcal{H}}_{\rm BF}&=U_{\rm BF}\sum_j
    \hat b^\dagger_{j}\hat f^\dagger_{j}\hat f_{j}\hat b_{j}, \label{eq:Hubbard_bf}
\end{align}
where $\hat b_{j}$ $(\hat b^\dagger_{j})$ and $\hat f_{j}$ $(\hat f^\dagger_{j})$ are the annihilation (creation) operators for bosons and fermions at the $j$-th site, respectively, with corresponding number operators $\hat n^{\rm B}_j = \hat b^\dagger_{j} \hat b_{j}$ and $\hat n^{\rm F}_j = \hat f^\dagger_{j} \hat f_{j}$. The hopping integral $t$ is assumed equal for both species, which has been realized in $^{171}\rm{Yb}-^{174}\rm{Yb}$~\cite{Takasu_JPSJ2009} and $^{6}\rm{Li}-^{7}\rm{Li}$~\cite{Ikemachi_JPB2016} experiments. The terms $U_{\rm BB}$ and $U_{\rm BF}$ parameterize the strength of the boson-boson and boson-fermion coupling respectively, tunable by the use of appropriate Feshbach resonances~\cite{Zaccanti_PRA2006, Best_PRL2009, Deh_PRA2010}. The BCs of each species are encoded in the hopping terms of their respective Hamiltonians \eqref{eq:Hubbard_b} and \eqref{eq:Hubbard_f}: for periodic BCs an additional hopping term between the last and first site of the chain is added while for open BCs that term is omitted. This type of mixed BCs can be engineered in experiments using a ring trap to enforce the periodic BCs with a potential barrier~\cite{Gallucci_NJP2016, Kumar_PRA2018, deGoerdeHerve_JPB2021} that breaks translational symmetry (open BCs) for only one species in the mixture employing species-dependent external potentials such as tuning-detuning schemes used for alkali atoms~\cite{LeBlanc_PRA2007}.

In the following, we fix the number of bosons at unit-filling, $N_{\rm B}=L$, with system size $L=30$, while the number of fermions is much smaller than the background bosons, $N_{\rm F} \ll N_{\rm B}$, for which we fix $N_{\rm F}=2$. Later we consider larger ratios of $N_{\rm F}/N_{\rm B}$ and system sizes. The interaction strengths will be varied keeping $t=1$ setting the energy scale. To explore the phase diagram of the coupled system \eqref{eq:Hubbard} the ground state is constructed using the Density Matrix Renormalization Group (DMRG) technique in the tensor network formalism~\cite{Schollwock_AP2011}. Details of our numerical simulations are given in Appendix~\ref{sec:DMRG}. 

\section{Polaron formation} \label{sec:formation}

\begin{figure}[tb]
    \centering
    \includegraphics[width=\linewidth]{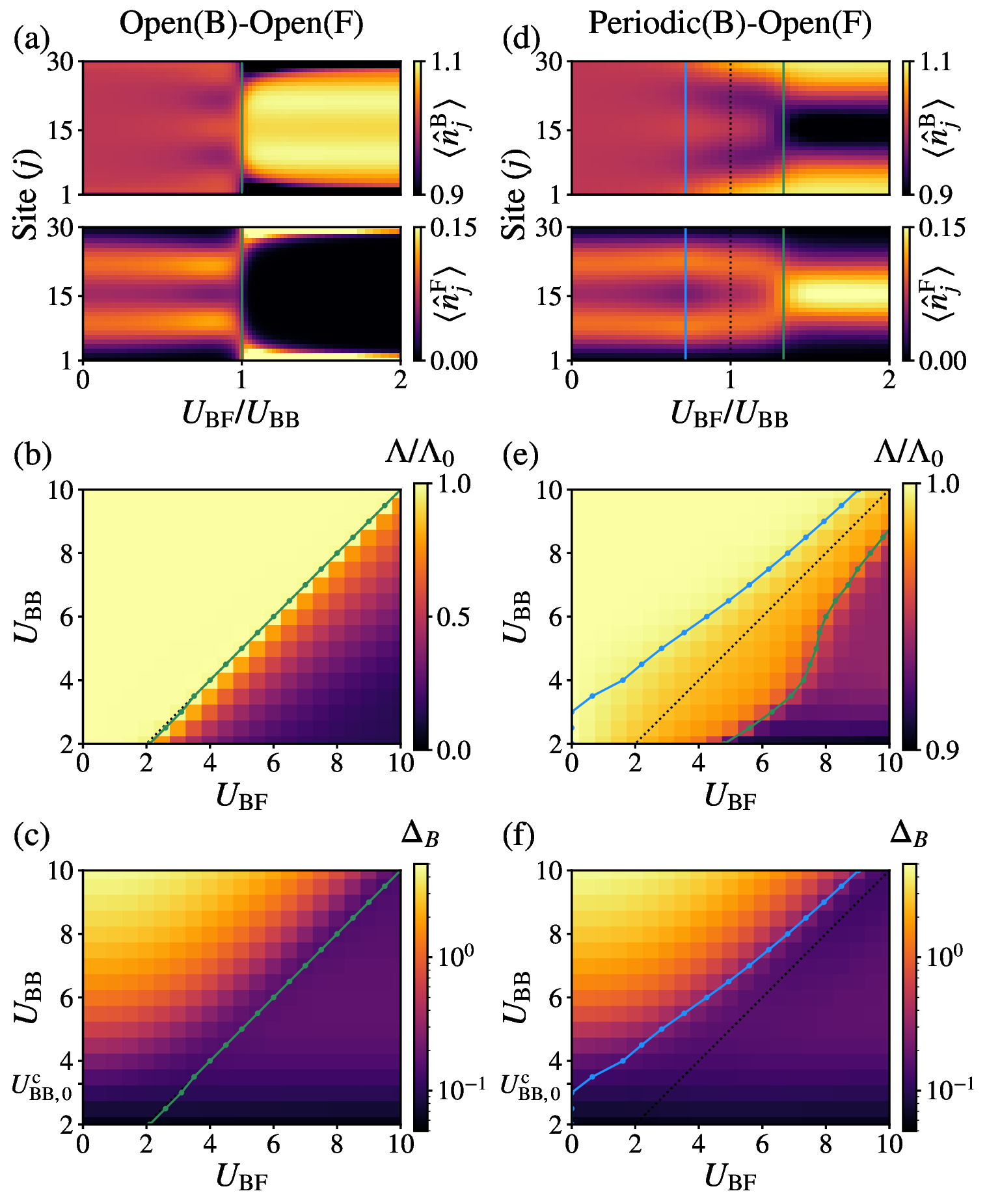}
    \caption{
        Local observables and bosonic energy gaps. The left and right rows correspond to Open(B)-Open(F) and Periodic(B)-Open(F) BCs, respectively.
        Panel (a) and (d): Density profile of the bosonic $\expval{\hat n^{\rm B}_j}$ and fermionic $\expval{\hat n^{\rm F}_j}$ species as a function of $U_{\rm{BF}}/U_{\rm{BB}}$ at $U_{\rm{BB}}=6$. Panel (b) and (e) show the normalized overlap $\Lambda/\Lambda_0$, and (c) and (f) the bosonic gap $\Delta_{\rm B}$ in logarithmic scale as a function of $U_{\rm{BF}}$ and $U_{\rm{BB}}$. Blue and green lines indicate the coupling strengths that give the maximum of the fermionic relative distance $D_{\rm FF}$ (see Fig.~\ref{fig:Distance}) and interspecies entropy $S_{\rm BF}$ (see Fig.~\ref{fig:CorrelationsOpenOpen},\ref{fig:CorrelationsClosedOpen}), respectively. The dotted black lines indicate $U_{\rm BB}=U_{\rm BF}$ for reference.
    } 
    \label{fig:DensityMiscibilityGap}
\end{figure}

\begin{figure}[tb]
    \centering
    \includegraphics[width=\linewidth]{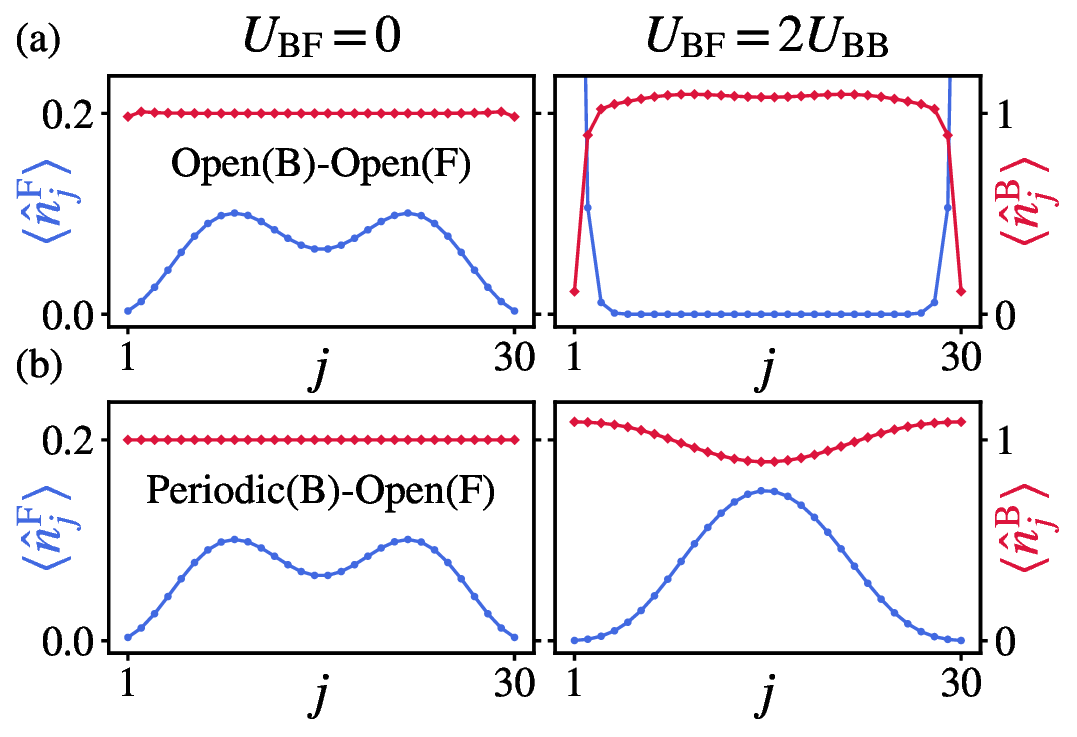}
    \caption{
        Density profile of the fermionic $\expval{\hat n^{\rm F}_j}$ (blue circles) and bosonic $\expval{\hat n^{\rm B}_j}$ (red diamonds) species at $U_{\rm{BB}}=6$ with (a) Open(B)-Open(F) (b) and Periodic(B)-Open(F) BCs for $U_{\rm{BF}}=0$ (left column) and $U_{\rm{BF}}/U_{\rm{BB}}=2$ (right column).
    }
    \label{fig:Density}
\end{figure}

We start by summarizing the case of Open(B)-Open(F) BCs and the well-known miscible-immiscible transition. 
In Fig.~\ref{fig:DensityMiscibilityGap}(a) we show the one-body density profiles of the mixture of the bosonic $\expval{\hat{n}^{\rm B}_j}$ and fermionic $\expval{\hat{n}^{\rm F}_j}$ species as a function of the interspecies coupling $U_{\rm BF}/U_{\rm BB}$ at fixed $U_{\rm BB} = 6$. 
In the decoupled system ($U_{\rm{BF}}=0$) exhibited in Fig.~\ref{fig:Density}(a), the background corresponds to a bosonic Mott insulator (MI) as the boson-boson interaction is much larger than the critical value of $U_{\rm{BB},0}^{\rm c}\approx3.28$~\cite{Kashurnikov_JETPL1996, Kuhner_PRB1998, Elstner_PRB1999, Kuhner_PRB2000, Ejima_EPL2011}
\footnote{Although this critical point is measured in the thermodynamic limit, it is still an important reference value for the transition in this finite-sized system.}, while the fermions are free and display two distinct peaks due to the Pauli repulsion.
For sufficiently small interspecies interaction, $U_{\rm BF}\ll U_{\rm BB}$, the rigid insulating nature of the MI prevents any notable distortion to both density profiles; however, the density profiles change dramatically when $U_{\rm BF}\geq U_{\rm BB}$ (see Fig.~\ref{fig:DensityMiscibilityGap}(a)) as the system becomes phase separated with the fermions being individually localized at the box edges, which we show explicitly for $U_{\rm{BF}}/U_{\rm BB}=2$ in Fig.~\ref{fig:Density}(a).

The miscible-immiscible transition can be captured by the interspecies density overlap $\Lambda$~\cite{Tylutki_NJP2016}
\begin{align}\label{eq:Overlap}
    \Lambda = \sum_j^{L} \expval{\hat{n}^{\rm B}_j} \expval{\hat{n}^{\rm F}_j},
\end{align}
shown in Fig.~\ref{fig:DensityMiscibilityGap}(b) as a function of $U_{\rm BF}$ and $U_{\rm BB}$. We scale the overlap by the corresponding decoupled value $\Lambda_{0}(U_{\rm BB}) \equiv \Lambda(U_{\rm BF}=0, U_{\rm BB})$ as a reference to the state of maximum miscibility for a given $U_{\rm BB}$.
The critical value is clear at $U_{\rm BB} \approx U_{\rm BF}$, with an abrupt decrease in $\Lambda$ for larger interspecies couplings, $U_{\rm BF} > U_{\rm BB}$. 
The emergence of phase separation necessitates that the filling fraction of the bosonic background must increase above unity around the center of the box as the fermions are localized at the box edges, resulting in the loss of the Mott phase. This is signaled by the closing of the energy gap 
\begin{align} \label{eq:Gap}
    \Delta_{\rm B} = E^L_{N_{\rm B}+1} -2E^L_{N_{\rm B}} + E^L_{N_{\rm B}-1},
\end{align}
where $E^L_{N_{\rm B}}$ is the ground state energy for $N_{\rm B}$ number of bosons with fixed $L$ number of lattice sites~\cite{Ejima_PRA2012}. 
In Fig.~\ref{fig:DensityMiscibilityGap}(c), we see that the gap reduces for increasing $U_{\rm BF}$, saturating to vanishingly small values in the phase separation regime $U_{\rm BF}>U_{\rm BB}$, indicating that a superfluid (SF) background is necessary for the miscible-immiscible transition to occur.

The nature of the transition alters dramatically if one considers Periodic(B)-Open(F) BCs, as seen from the density profiles in Fig.~\ref{fig:DensityMiscibilityGap}(d). 
In particular, the most relevant change occurs for $U_{\rm BF} \gtrapprox U_{\rm BB}$ where the absence of a boundary effect for the bosonic cloud (compare edges of the bosonic profile between BCs in Fig.~\ref{fig:Density} at $U_{\rm BF}=0$) prevents the system from transitioning to an immiscible state. 
Instead, the fermions cluster together and reorganize themselves inside the bulk of the lattice on top of a reduced but relatively large bosonic accumulation shown in Fig.~\ref{fig:Density}(b) for $U_{\rm BF}/U_{\rm BB}=2$. 

The absence of the phase separated state can be captured by $\Lambda$ (see Fig.~\ref{fig:DensityMiscibilityGap}(e)), where for fixed $U_{\rm BB}$ and increasing $U_{\rm BF}$, the overlap decreases smoothly rather than the abrupt transition in the Open(B)-Open(F) BCs case, and saturates at a finite and relatively large value. Indeed, the density of the fermions does not change with increasingly large boson-fermion coupling $U_{BF}/U_{BB}\gg 1$, with the fermions remaining spread over many lattice sites and any further localization being suppressed. In this regime the fermions have formed a tightly bound bipolaron state due to the strong mediated attractive interactions from the boson bath, which we will further characterize in the following sections.

Finally, as in the case of the miscible-immiscible transition, the SF bosonic background is necessary for this re-ordering of the fermions to occur, as indicated in Fig.~\ref{fig:DensityMiscibilityGap}(f), where $\Delta_{\rm B}$ behaves similarly to the Open(B)-Open(F) BCs case (Fig.~\ref{fig:DensityMiscibilityGap}(c)) since this is a bulk property. 

\par%

\section{SF-MI transition and fermionic relative distance} \label{sec:SF_MI}

\begin{figure}[h]
    \centering
    \includegraphics[width=\linewidth]{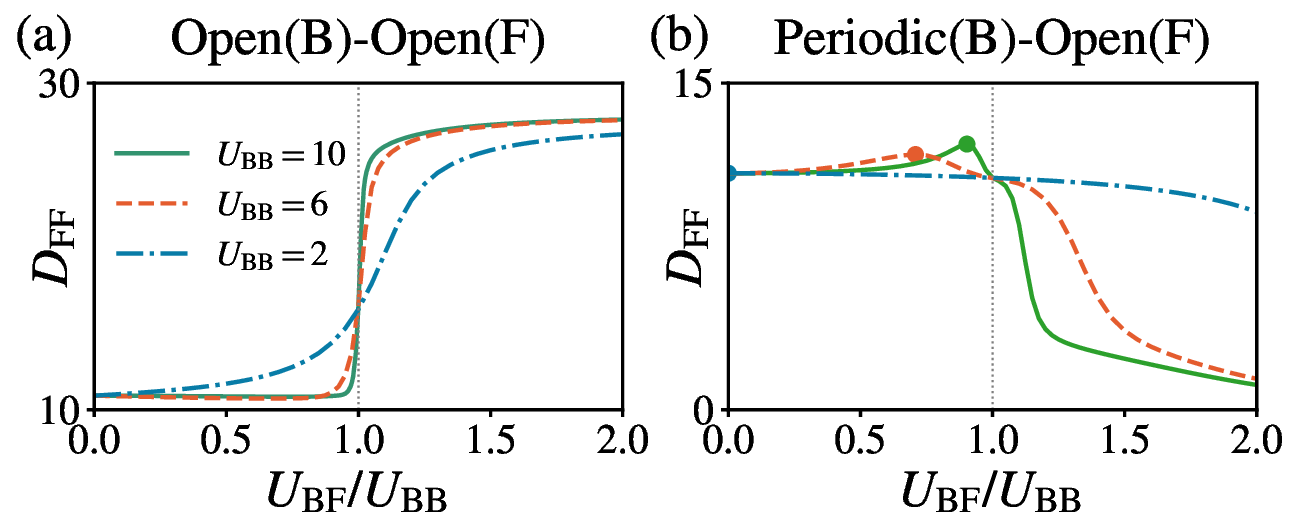}
    \caption{Fermionic relative distance $D_{\rm FF}$ as a function of $U_{\rm BF}/U_{\rm BB}$ for different $U_{\rm BB}$ values and with (a) Open(B)-Open(F) and (b) Periodic(B)-Open(F) BCs. The dot on each curve in panel (b) indicates the maximum distance for corresponding $U_{\rm BB}$. The dotted line references $U_{\rm BF}=U_{\rm BB}$.}
    \label{fig:Distance}
\end{figure}

Besides the polaron formation, the Periodic(B)-Open(F) BCs exhibit an intrinsic relation between the SF-MI transition of the bosonic bath and the distribution of the fermionic impurities. This can be succinctly seen in the relative distance between the two fermions~\cite{Mistakidis_PRA2019}
\begin{align} \label{eq:Distance}
    D_{\rm FF} = \frac{1}{N_{\rm F}(N_{\rm F}-1)} \sum_{j,k}^L |k-j| \expval{\hat f^\dagger_j \hat f^\dagger_k \hat f_k \hat f_j}\,,
\end{align}
 shown in Fig.~\ref{fig:Distance} as a function of $U_{\rm BF}/U_{\rm BB}$ for different $U_{\rm BB}$ values and BCs.
As a reference, we first analyze the case for Open(B)-Open(F) BCs (Fig.~\ref{fig:Distance}(a)). Here the fermions appear to spread along the lattice progressively as we increase $U_{\rm BF}$ which agrees with the behavior at the miscible-immiscible transition ($U_{\rm BF} \approx U_{\rm BB}$) where the relative distance rapidly increases indicating the separation of the impurities. This increase is more abrupt for higher $U_{\rm BB}$ due to the breaking of the MI phase via the interaction with the fermions. 

Periodic(B)-Open(F) BCs (Fig.~\ref{fig:Distance}(b)) present a different behavior for $U_{\rm BB}$ values below and above $U_{\rm BB, 0}^{\rm c}$. In the former ($U_{\rm BB} = 2$ in Fig.~\ref{fig:Distance}(b)), $D_{\rm FF}$ decreases as the fermions grow closer with increasing $U_{\rm BF}$ as a prelude of the polaron formation. On the other hand, for $U_{\rm BB} > U_{\rm BB, 0}^{\rm c}$ ($U_{\rm BB} = 6$, $10$ in Fig.~\ref{fig:Distance}(b)) before the polaron emergence at $U_{\rm BF}/U_{\rm BB} \approx 1$ there is an increase in $D_{\rm FF}$.
% which cannot be associated with the bound state formation.
Here as the background is in the MI phase, bosonic correlations are short-ranged~\cite{Bellomia_PRB2024}, and therefore bosons only couple locally to the fermions, leading to weak self-localization of individual impurities that can be seen as an increase in the fermionic density maxima and the emergence of minima in the bosonic density in Fig.~\ref{fig:DensityMiscibilityGap}(d). After the SF-MI transition, long-range correlations that characterize the SF bath can, in turn, mediate interactions between the impurities longer than the MI bath which leads them to cluster together around the center of the lattice, decreasing $D_{\rm FF}$ as it is for $U_{\rm BB} < U_{\rm BB, 0}^{\rm c}$. It is important to note that the fermionic self-localization effect occurs for both BCs, given that this is a phenomenon of the MI bulk, and is also seen in Fig.~\ref{fig:DensityMiscibilityGap}(a) for $U_{\rm BF}/U_{\rm BB}<1$. 

% Since this change in the monotonic behavior of the $D_{\rm FF}$ is directly related to the bosonic background's phase, a maximum in $D_{\rm FF}$ signals the SF-MI transition, shown as a dot on each curve in Fig.~\ref{fig:Distance}(b). 
Therefore, a characteristic maximum in $D_{\rm FF}$ appears when the impurities drive the SF-MI transition of the bosonic background's phase, which is marked by a dot on each curve in Fig.~\ref{fig:Distance}(b). 
We highlight this maximum relative distance for each $U_{\rm BF}$ and $U_{\rm BB}$ with solid blue lines in Fig.~\ref{fig:DensityMiscibilityGap}(d-f). Remarkably, this clearly delineates the SF-MI phase boundary by agreeing with the other phase transition indicators, like the breaking of the bosonic uniform density profile $\expval{\hat{n}_j^{\rm B}}$ in Fig.~\ref{fig:DensityMiscibilityGap}(d), the decrease from maximum miscibility in the overlap $\Lambda$ in Fig.~\ref{fig:DensityMiscibilityGap}(e) and the vanishing of the energy gap $\Delta_B$ in Fig.~\ref{fig:DensityMiscibilityGap}(f).
Moreover, we observe that by increasing $U_{\rm BB}$ the maximum of the relative distance in Fig.~\ref{fig:Distance}(b) also increases. This occurs because $\Delta_{\rm B}$ grows with $U_{\rm BB}$ as shown in Fig.~\ref{fig:DensityMiscibilityGap}(f), hence the MI phase becomes more stable against the perturbation by the impurities, which in turn, allows the impurities to self-localize further without breaking the MI state. We provide further characterization of the bosonic correlations in the transition in Appendix~\ref{sec:SF_MI_corr}.

%It is important to note that the fermionic self-localization effect occurs for both BCs, given that this is a phenomenon of the bulk, as seen in Fig.~\ref{fig:DensityMiscibilityGap}(a,d) for $U_{\rm BF}/U_{\rm BB}<1$. 
%Still, for Open(B)-Open(F) BCs $D_{\rm FF}$ is a monotonic increasing function of $U_{\rm BF}/U_{\rm BB}$ for a fixed $U_{\rm BB}$ given that phase separation further separates the impurities, therefore there is no change in the sign of the derivative of the curve that can indicate the SF-MI transition and there is no other evident effect in the density profiles that we know of.

\section{Strong coupling regime} \label{sec:strong_regime}

To further characterize the state of the fermions in the regime of strong impurity-bath coupling we calculate its one- and two-body reduced density matrices given by 
\begin{eqnarray}
    \hat \rho_{\rm F}^{(1)} &=& \frac{1}{N_{\rm F}} \expval{ \hat f^\dagger_j \hat f_k} ,\label{eq:OneBody}\\  
     \hat \rho_{\rm F}^{(2)} &=& \frac{1}{N_{\rm F}(N_{\rm F}-1)}  \expval{\hat f^\dagger_j \hat f^\dagger_k \hat f_m \hat f_l} , \label{eq:TwoBody} 
\end{eqnarray} 
respectively. The one-body reduced density matrix is an auto-correlation function describing the probability of finding the same fermion simultaneously in two different lattice sites $j$ and $k$, and we have already studied the contribution of its diagonal ($j=k$) in the form of fermionic density profiles $\expval{\hat{n}^{\rm F}_j}$ in Fig.~\ref{fig:DensityMiscibilityGap}. On the other hand, the two-body reduced density matrix describes the joint probability of finding one fermion in sites $j$ and $l$ and the other fermion in sites $k$ and $m$, which was used to define $D_{\rm FF}$ in \eqref{eq:Distance}. As $\hat \rho_{\rm F}^{(2)}$ is difficult to visualize we instead study its diagonal contribution i.e. when $l=j$ and $m=k$, via the two-body density profile
\begin{align}
    \expval{\hat{n}^{\rm FF}_{j,k}} 
    &= \expval{\hat f^\dagger_j \hat f^\dagger_k \hat f_k \hat f_j} 
    % &= \expval{\hat f^\dagger_j \hat f^\dagger_k (- \hat f_j \hat f_k)}, \\
    = \expval{\hat f^\dagger_j (\hat f_j \hat f^\dagger_k - \delta_{j,k}) \hat f_k} \notag\\
    % &= \expval{\hat n^{\rm F}_j \hat n^{\rm F}_k} - \delta_{j,k} \expval{\hat n^{\rm F}_j \hat n^{\rm F}_j}, \\
    &=(1-\delta_{j,k}) \expval{\hat{n}^{\rm F}_j \hat{n}^{\rm F}_k},
\end{align}
where we used the anticommutation properties of fermionic operators along with $\hat n^{\rm F}_j(\hat n^{\rm F}_j-1) = 0$.
This represents the probability of finding one fermion at site $j$ and the other at site $k$, and is, of course, zero at the diagonal ($j=k$) due to the Pauli exclusion principle. 

\begin{figure}[tb]
    \centering
    \includegraphics[width=\linewidth]{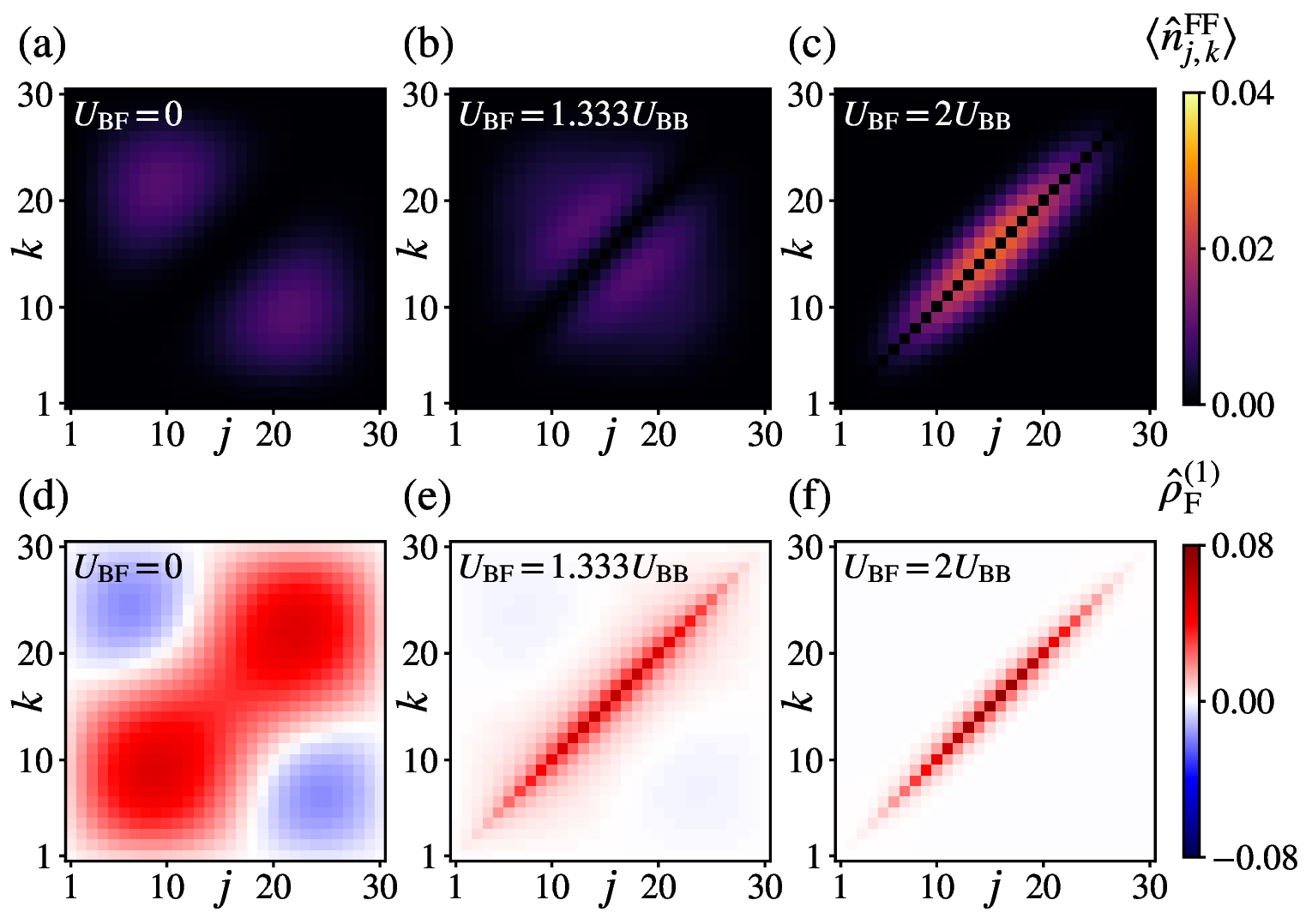}
    \caption{ (a-c) Two-body fermionic density profiles $\langle \hat n^{\rm{FF}}_{j,k}\rangle$ and (d-f) one-body fermionic reduced density matrix $\hat \rho_{\rm F}^{(1)}$ for Periodic(B)-Open(F) BCs. Here we have fixed the boson-boson coupling $U_{\rm BB}=6$ and take (a,d) $U_{\rm BF}=0$, (b,e) $U_{\rm BF}=1.333\,U_{\rm BB}$, and (c,f) $U_{\rm BF}=2\,U_{\rm BB}$.
    }
    \label{fig:TwoBodyMeasures}
\end{figure}

We start with an analysis of the two-body density profile shown in Fig.~\ref{fig:TwoBodyMeasures}(a-c) for $U_{\rm BB} = 6$. Free fermions are anti-correlated (Fig.~\ref{fig:TwoBodyMeasures}(a) for $U_{\rm BF}=0$) as the two particles are separated into two different sides of the box potential. 
However, for finite interspecies coupling $U_{\rm BF}=1.333 U_{\rm BB}$ (Fig.~\ref{fig:TwoBodyMeasures}(b)), which is after the MI-SF transition for the bosons and they are therefore in the SF phase, the fermions start to bunch (they are more likely to be found near one another) and the relative distance between them decreases (see also Fig.~\ref{fig:Distance}(b)).
For even larger $U_{\rm BF}$ (Fig.~\ref{fig:TwoBodyMeasures}(c)) the fermions form a tightly bound state and $\langle \hat n^{\rm{FF}}_{j,k}\rangle$ is significantly broadened along the $j=k$ direction, indicating a large variance in the center of mass position (they are spread across the lattice) and enhanced particle bunching (the fermions are most likely to be found in adjacent lattice sites).
A similar behaviour can be seen in the one-body density matrix shown in Fig.~\ref{fig:TwoBodyMeasures}(d-f), with the tightly bound state only displaying short-range off-diagonal coherences in Fig.~\ref{fig:TwoBodyMeasures}(f). The emergence of this tightly bound fermionic state is due to effective attractive interactions that are mediated by the coupling to the bosonic background and are indicative of polaron physics~\cite{Fritsche_PRA2021, Baroni_NatP2024}, confirming that a bipolaron state is formed when considering Periodic(B)-Open(F) BCs.

% A key concept in the analysis of polaron physics is the quasiparticle weight~\cite{Grusdt_2024}, defined as the squared overlap between the interacting wave function $\ket{U_{\rm BF}, U_{\rm BB}}$ with the decoupled state $\ket{U_{\rm BF}=0, U_{\rm BB}}$:
% \begin{align}
%     Z=|\braket{U_{\rm BF}=0, U_{\rm BB}}{U_{\rm BF},U_{\rm BB}}|^2.
% \end{align}
% In Fig.~\ref{fig:QuasiparticleWeight} we show this quantity as a function of the interspecies interaction for multiple $U_{\rm BB}$ values, all of which show the convergence of $Z\to 0$ as we increase $U_{\rm BF}$. This indicates that the $N$-polaron state observed in this work falls outside the quasiparticle picture as the coupled bath-impurities system changes drastically from the non-interacting case, which can be observed in Fig.~\ref{fig:Density}(b).
% \begin{figure}[tb]
%     \centering
%     \includegraphics[width=0.9\linewidth]{QuasiparticleWeight.eps}
%     \caption{ Quasiparticle weight $Z$ as a function of the interspecies interaction $U_{\rm BF}$ scaled by a fixed $U_{\rm BB} = 2$, $4$ and $6$. $Z=0$ is shown as a dotted line for reference. 
%     } 
%     \label{fig:QuasiparticleWeight}
% \end{figure}

\par%

The bipolaron state emerges when the coupling to the bosonic background is large $U_{\rm BF}\gg U_{\rm BB}$ and there is significant overlap between bosons and fermions $\Lambda/\Lambda_0\gg0$ (see Fig.~\ref{fig:DensityMiscibilityGap}(e)). This implies that there are strong interspecies correlations between the two components and that these play an important role in polaron formation. The boson-fermion correlation can be quantified through the von Neumann entropy of the fermionic reduced density matrix, $S_{\rm BF}=-\Tr \left[ \hat \rho_{\rm F} \ln{\hat \rho_{\rm F}} \right]$, where $\hat \rho_{\rm F}=\Tr_{\rm B} \hat \rho$ is the reduced fermionic state after tracing out the bosonic component. For $N_{\rm F}=2$, which we consider here, $\hat \rho_{\rm F}$ is given by the two-body reduced density matrix $\hat \rho_{\rm F}^{(2)}$ in Eq.~\eqref{eq:TwoBody}. Similarly, the correlations between the two fermions can be quantified through the mutual information $I_{\rm FF}= 2S^{(1)}_{\rm F} - S_{\rm BF}$, with $S^{(1)}_{\rm F}$ the von Neumann entropy of the one-body reduced density matrix given by Eq.~\eqref{eq:OneBody}. 

\par%

\begin{figure}[tb]
    \centering
    \includegraphics[width=0.9\linewidth]{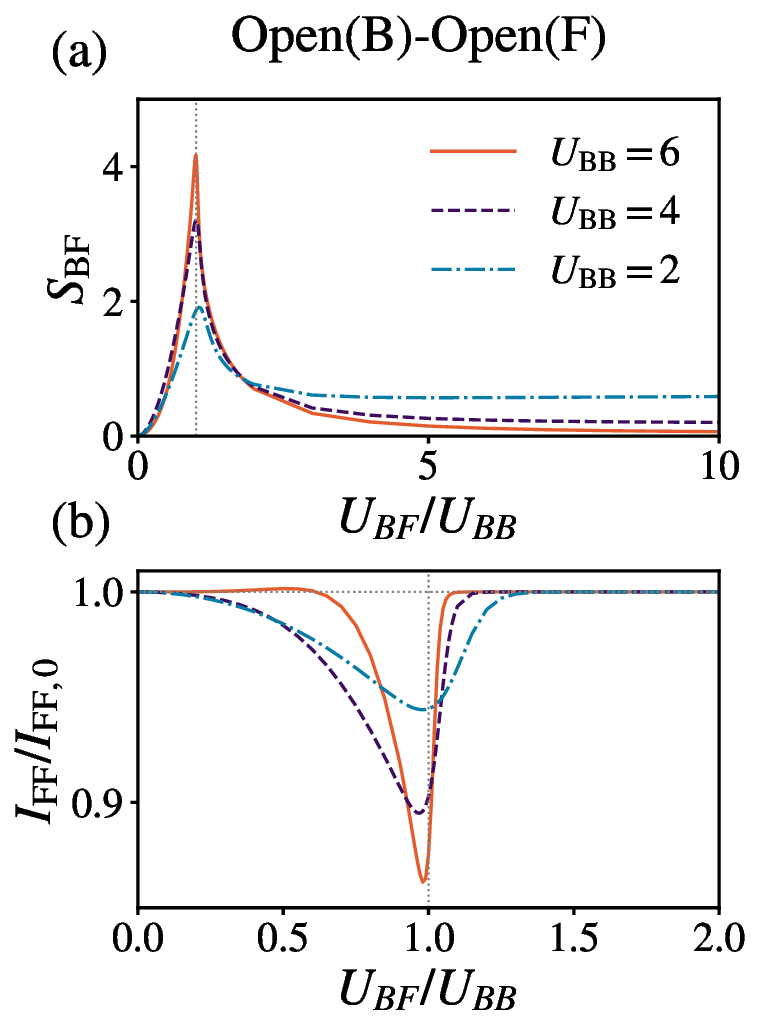}
    \caption{
        (a) Interspecies entropy $S_{\rm BF}$ and (b) fermionic mutual information $I_{\rm FF}$ as a function of $U_{\rm BF}/U_{\rm BB}$ for $N_{\rm F}=2$, fixed $U_{\rm BB} = 2$ (blue dot-dashed), $4$ (purple dashed) and $6$ (orange solid) and Open(B)-Open(F) BCs. 
        $I_{\rm FF}$ is normalized to the non-interacting value $I_{\rm{FF},0}=I_{\rm FF}(U_{\rm BF}=0)$. Gray dotted lines indicate $U_{\rm BF}= U_{\rm BB}$ (vertical) and $I_{\rm FF} = I_{\rm{FF}, 0}$ (horizontal) for reference. 
    } 
    \label{fig:CorrelationsOpenOpen}
\end{figure}
\begin{figure}[tb]
    \centering
    \includegraphics[width=0.9\linewidth]{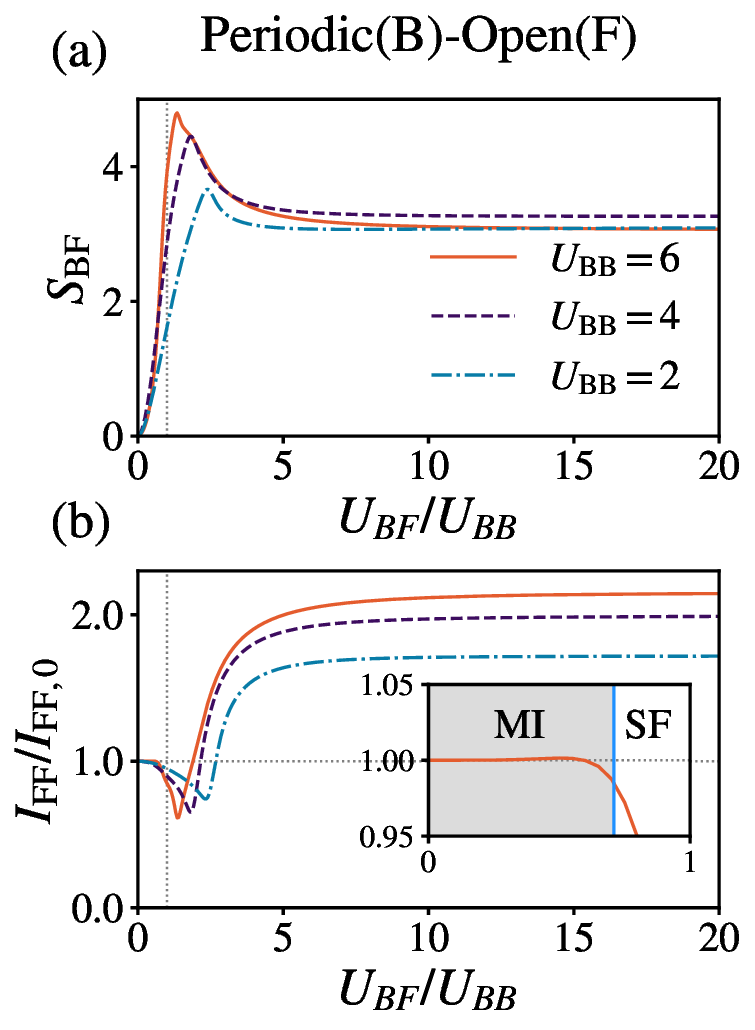}
    \caption{
        (a) Interspecies entropy $S_{\rm BF}$ and (b) fermionic mutual information $I_{\rm FF}$ as a function of $U_{\rm BF}/U_{\rm BB}$ for $N_{\rm F}=2$, fixed $U_{\rm BB} = 2$ (blue dot-dashed), $4$ (purple dashed) and $6$ (orange solid) and Periodic(B)-Open(F) BCs. Inset in (b): Closeup of the mutual information across the SF-MI transition with the blue line obtained from Fig.~\ref{fig:Distance}.
        $I_{\rm FF}$ is normalized to the non-interacting value $I_{\rm{FF},0}=I_{\rm FF}(U_{\rm BF}=0)$. Gray dotted lines indicate $U_{\rm BF}= U_{\rm BB}$ (vertical) and $I_{\rm FF} = I_{\rm{FF}, 0}$ (horizontal) for reference. 
    } 
    \label{fig:CorrelationsClosedOpen}
\end{figure}
We first explore the case of Open(B)-Open(F) BCs where phase separation occurs, for which we show the von Neumann entropy and mutual information in Fig.~\ref{fig:CorrelationsOpenOpen}(a) and (b) as a function of $U_{\rm BF}$.
Starting from the decoupled pure states at $U_{\rm BF}=0$, $S_{\rm BF}$ increases with increasing $U_{\rm BF}$ as bosonic and fermionic states mix, reaching its maximum around the critical interaction strength $U_{\rm BF}\approx U_{\rm BB}$.
In Fig.~\ref{fig:DensityMiscibilityGap}(a-c) we indicate the coupling strengths that give the maximum entropy with green lines, showing that it exactly separates the two phases and only shows very small deviations from $U_{\rm BF}=U_{\rm BB}$ below $U_{\rm BB,0}^c$. 
On the contrary, the mutual information $I_{\rm FF}$  decreases until the miscible-immiscible transition occurs, indicating impurity-impurity correlation screening due to the increased
correlations with the bath, reaching a minimum near the
$S_{\rm BF}$ maximum~\footnote{For small $U_{\rm BB}$, the $S_{\rm BF}$ and $I_{\rm FF}$ extrema slightly differ but coincide as $U_{\rm BB}$ increases.}. Increasing $U_{\rm BF}$ beyond this critical point leads to the fermions moving the systems edges and a decay in $S_{\rm BF}$ while $I_{\rm FF}$ returns to its initial value.
These are both indicators of the phase separation regime in which the only contribution to the impurity-impurity correlations is the Pauli exclusion principle, while the bosonic bath acts as an effective double well potential that confines the two impurities to the trap edges. The residual correlation in $S_{\rm BF}$ at high $U_{\rm BF}$ values is larger when the system is deeper in the SF regime ($U_{\rm BB}<U_{\rm BB,0}^c$), which can be understood as the compressible nature of the bosonic SF allowing a finite overlap with the fermions even at large $U_{\rm BF}$.

\par

%For Periodic(B)-Open(F) BCs with weak interspecies coupling, $U_{\rm BF}<U_{\rm BB}$, $S_{\rm BF}$ and $I_{\rm FF}$ exhibit similar behavior as for Open(B)-Open(F) BCs, increasing and decreasing respectively until reaching a critical point that signals the miscible-polaronic transition. Moreover, we note that there is a plateau at $I_{\rm FF} \approx I_{\rm{FF}, 0}$ in the region $U_{\rm BF} < U_{\rm BB}$ for both BCs with $U_{\rm BB}=6$ (Fig.~\ref{fig:Correlations}(b,d)) which is an indicator of the MI background where additional correlations are suppressed~\cite{Bellomia_PRB2024}, as discussed in Sec.~\ref{sec:SF_MI}. \tf{i think we need the inset back. We have free space now so we can just make the figure bigger to accommodate it. I think the referee was mostly compaing about the other inset anyway....}

%Again we take the coupling strength at the maximum of the von Neumann entropy and highlight it by the green lines in Fig.~\ref{fig:DensityMiscibilityGap}(d) and (e), which clearly signals the emergence of the bipolaron state. 

\par%

For Periodic(B)-Open(F) BCs (Fig.~\ref{fig:CorrelationsClosedOpen}) with weak interspecies coupling, $U_{\rm BF}<U_{\rm BB}$, $S_{\rm BF}$ and $I_{\rm FF}$ exhibit similar behavior as for Open(B)-Open(F) BCs, increasing and decreasing respectively until reaching a critical point that signals the miscible-polaronic transition. Again we take the coupling strength at the maximum of the von Neumann entropy and highlight it by the green lines in Fig.~\ref{fig:DensityMiscibilityGap}(d) and (e), which clearly signals the emergence of the bipolaron state. Moreover, in the inset of Fig.~\ref{fig:CorrelationsClosedOpen}(b), we show in a closeup of the mutual information for $U_{\rm BB}=6$ that while in the MI phase (filled by grey background) $I_{\rm FF}$ is unchanged, indicating a lack of impurity-impurity correlations due to the ``slow" spread of information through the insulating background~\cite{Bellomia_PRB2024}, until the bath transitions to the SF state (white background).
This transition almost coincides with the critical coupling strength obtained by the maximum relative distance between the impurities indicated by the blue vertical line, which is highlighted previously in Fig.~\ref{fig:DensityMiscibilityGap}(d-f).

\begin{figure}[h]
    \centering
    \includegraphics[width=\linewidth]{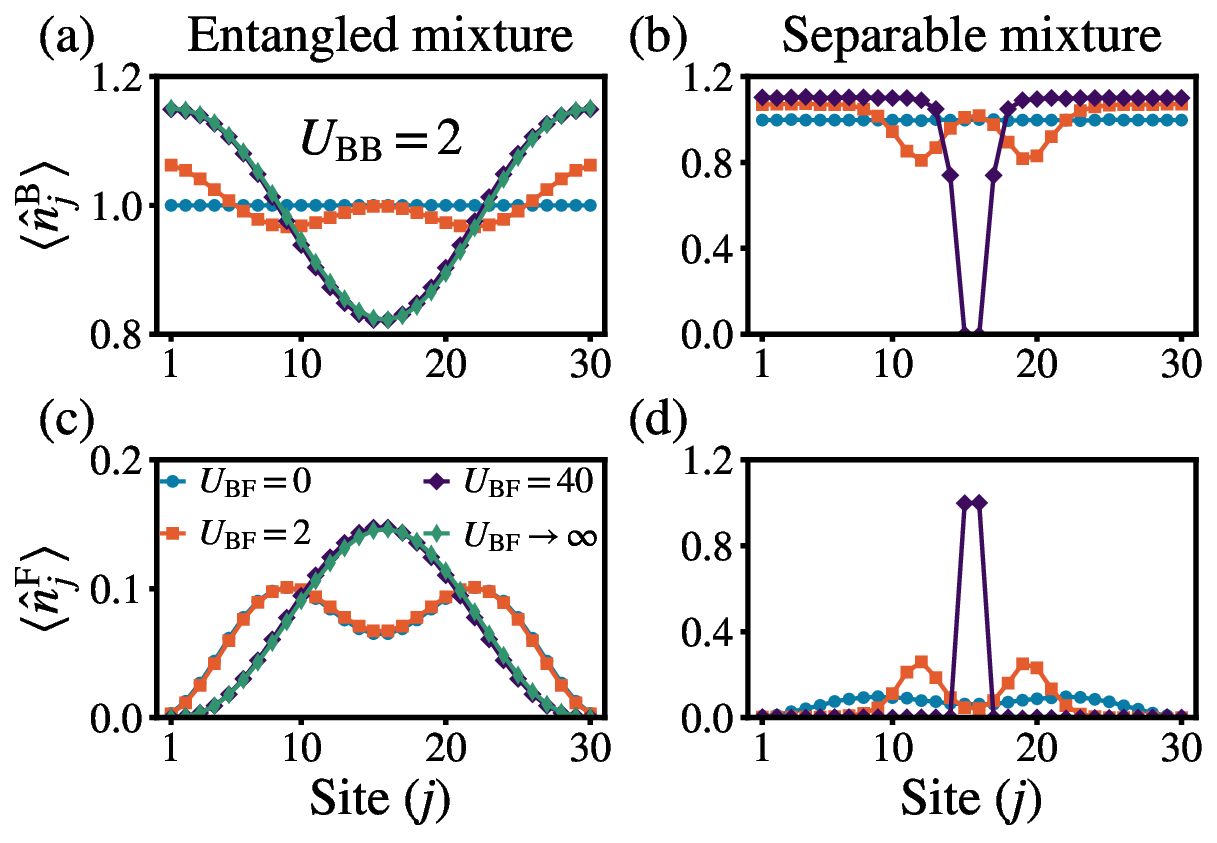}
    \caption{Comparison of density profiles between a general entangled mixture ansatz (a,c) and a separable (product) ansatz (b,d). (a,b) Bosonic $\expval{\hat{n}^{\rm B}_j}$ and (c,d) fermionic $\expval{\hat{n}^{\rm F}_j}$ density profiles for fixed $U_{\rm BB}=2$, multiple $U_{\rm BF}$ values interactions, including the $U_{\rm BF} \to \infty$ limit for the entangled mixture, and with Periodic(B)-Open(F) BCs.
    }
    \label{fig:DensitiesEntanglement}
\end{figure}

As expected from the differences in density profiles (Fig.~\ref{fig:DensityMiscibilityGap}(a,d)), the behavior changes radically after the miscible-polaron transition occurs where we observe a notable increase in $I_{\rm FF}$ and $S_{\rm BF}$ until both quantities saturate to large values in the $U_{\rm BF} \to \infty$ limit. Indeed, this indicates that strong impurity-bath correlations are responsible for the increased impurity-impurity correlations, as the former is necessary to induce large attractive interactions between the impurities and thus create the tightly bound states shown in Fig.~\ref{fig:TwoBodyMeasures}(c). To confirm the crucial role of correlations in the formation of the polaron state we compare with a Bose-Fermi mixture which is separable and therefore uncorrelated. In Fig.~\ref{fig:DensitiesEntanglement}(a,c) we show the bosonic and fermionic densities with interspecies correlations, and the emergence of the polaron state for increasing $U_{BF}$. We find that for strong coupling $U_{BF}=40$ the density converges to the $U_{\rm BF} \to \infty$ limit, the latter which we obtain by removing the double Bose--Fermi occupancy in the Hilbert space for the DMRG calculations (see Appendix~\ref{sec:DMRG}). 

In comparison, we show in Fig.~\ref{fig:DensitiesEntanglement}(b,d) the densities for the mixture without interspecies correlations, by using a separable state ansatz where we employ a mixed DMRG-exact diagonalization approach for the ground state search (see Appendix~\ref{sec:DMRG}). Here we observe the direct effect of the density-density coupling which confines the fermions in a deeper effective potential which is formed by the bosonic density minima in the centre of the system ($U_{\rm BF}=2$). For large interspecies interactions ($U_{\rm BF}=40$) the system reaches phase separation with the fermions completely localized on two adjacent lattice sites at the center of the trap surrounded by the bosonic density which is entirely depleted at the position of the fermions. At such large coupling, the bosonic translational symmetry is spontaneously broken with an $L$-fold degenerate ground-state labeled by the position of the bipolaron, as suggested by previous works on mixtures~\cite{Schonmeier-Kromer_PRB2023}. It is important to note that without impurity bath-correlations there are no induced interactions between the fermions ($I_{\rm FF}=I_{\rm{FF},0}$), with their apparent attraction and self-localization being wholly due to the local bosonic effective potential.

%Given that we obtained this state only by removing the interspecies correlations, we show that these block the formation of the phase separation and give rise to the polaron structure in discrete lattice systems analog to results in Ref.~\cite{Zschetzsche_PRR2024} for continuous systems using a variational ansatz in which the impurity-bath correlations are described nonperturbatively.

As the phase-separated regime in Fig.~\ref{fig:DensitiesEntanglement}(b,d) can only be reached by neglecting interspecies correlations, it highlights the important role that strong impurity-bath entanglement plays in polaron formation and impurity self-localization. This is typified by the saturation of the von Neumann entropy in Fig.~\ref{fig:CorrelationsClosedOpen}(a) which signifies how correlations are locked between the impurities and the bath in the strong coupling regime, and therefore halts further localization of the fermions as shown in Fig.~\ref{fig:DensityMiscibilityGap}(d). This is an example of impurity self-localization suppression enhanced by impurity-bath correlations, an effect which was proposed in Ref.~\cite{Zschetzsche_PRR2024} using a variational ansatz in which the impurity-bath correlations are described nonperturbatively. Our work provides further evidence for this effect in discrete lattice systems, and here it is also responsible for the system to completely avoid the phase-separated state.

\section{Role of population imbalance}\label{sec:populationimbalance}

\begin{figure}[tb]
    \centering
    \includegraphics[width=\linewidth]{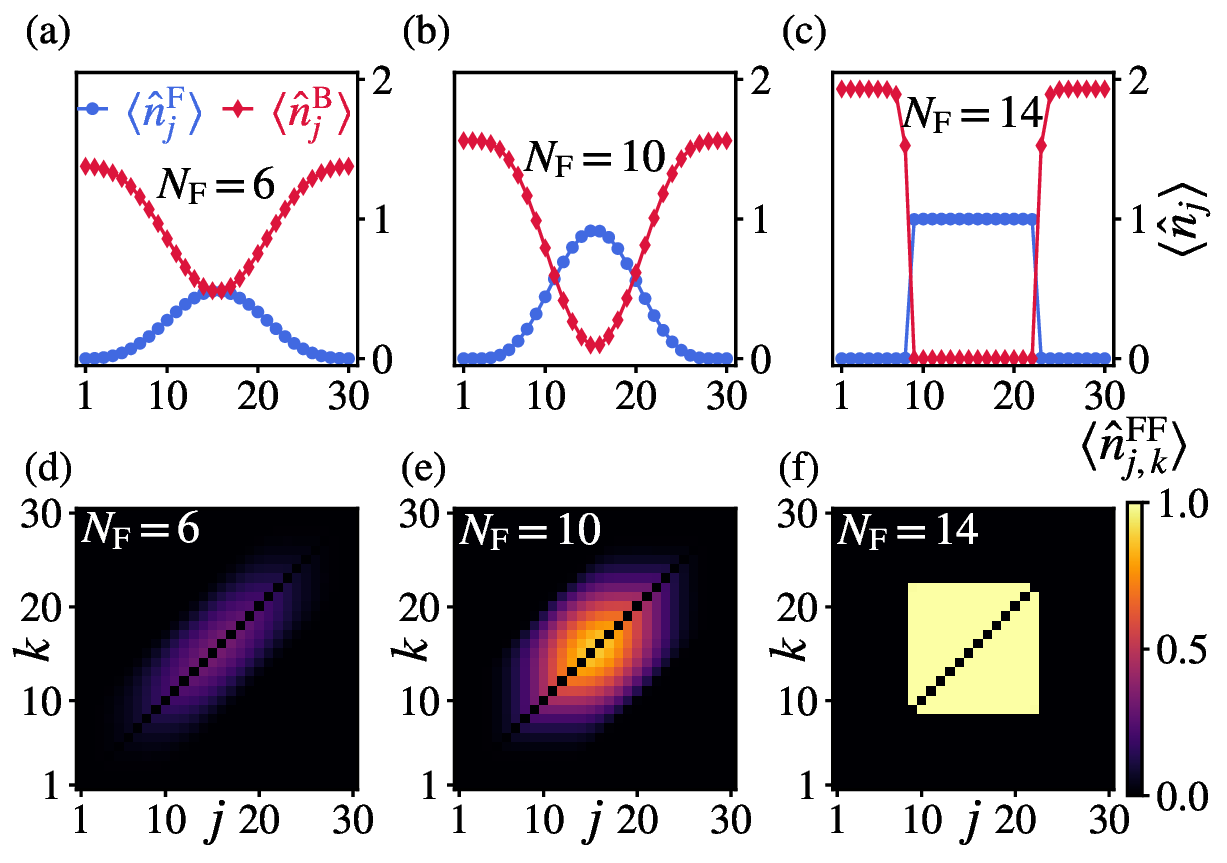}
    \caption{ 
        (a,b,c) One-body density profiles for fermions (blue circles) and bosons (red diamonds) with different numbers of impurities at $U_{\rm BB}=2$ and $U_{\rm BF} \to \infty$. (d,e,f) Two-body fermionic density profiles at the same interaction parameters and corresponding impurity numbers.
    } 
    \label{fig:PopulationImbalance}
\end{figure}

Next, we investigate systems with larger numbers of fermions, $N_{\rm F} > 2$, while keeping the number of bosons fixed at $N_{\rm B}=L=30$. In Fig.~\ref{fig:PopulationImbalance}(a-c) we show the density profiles for fixed interactions $U_{\rm BB}=2$ and $U_{\rm BF} \to \infty$, for impurity sizes $N_{\rm F}= 6$, $10$ and $14$. 
We see in Fig.~\ref{fig:PopulationImbalance}(a,b), for $N_{\rm F}=6$ and $N_{\rm F}=10$ respectively, that the ground states are $N$-polaron states with density profiles of similar characteristic shape as that in Fig.~\ref{fig:Density}(b). We also note that the fermions become more localized in the centre of the trap as the number of impurities is increased, with the bosonic density consequently decreasing in the same region. For larger numbers of impurities, $N_{\rm F}=14$ in Fig.~\ref{fig:PopulationImbalance}(c), there is no signature of the $N$-polaron state and instead the system is phase separated with the bosonic component is completely depleted at the centre of the trap while the fermions are at unit filling. This shows that the formation of the $N$-polaron state is only present when there is a high interspecies imbalance. 

In general, adding impurities to the system at fixed $U_{\rm BB}$ effectively compresses the bath, as the bosons cannot occupy the same place as the fermionic cloud, which is explicitly shown in the bosonic profiles through Fig.~\ref{fig:PopulationImbalance}(a-c). With a denser background, bath-mediated interactions are suppressed which removes the possibility of delocalization of the fermionic cluster and precipitates the collapse of the polaronic state. We can see this in the corresponding two-body density profiles in Fig.~\ref{fig:PopulationImbalance}(d-f), which for $N_{\rm F}=6$ and $N_{\rm F}=10$ (Fig.~\ref{fig:PopulationImbalance}(d,e)) shows qualitatively similar features as in Fig.~\ref{fig:TwoBodyMeasures}(c). However, it is apparent that the fermions are progressively localized along the center of mass direction ($i=j$) and delocalized along the relative coordinate direction ($i=-j$) as the number of impurities increases and the mediated interactions are diminished. In the phase separation regime at $N_{\rm F} = 14$ ((Fig.~\ref{fig:PopulationImbalance}(f)) the two-body density is uniform with equal probability to find a fermion anywhere in the central $14$ lattice sites, with all signature of the polaron state lost.
%, compacted density with a center of mass completely localized in the center of the system which represents the phase-separated state in this two-body picture.

\begin{figure}[tb]
    \centering
    \includegraphics[width=\linewidth]{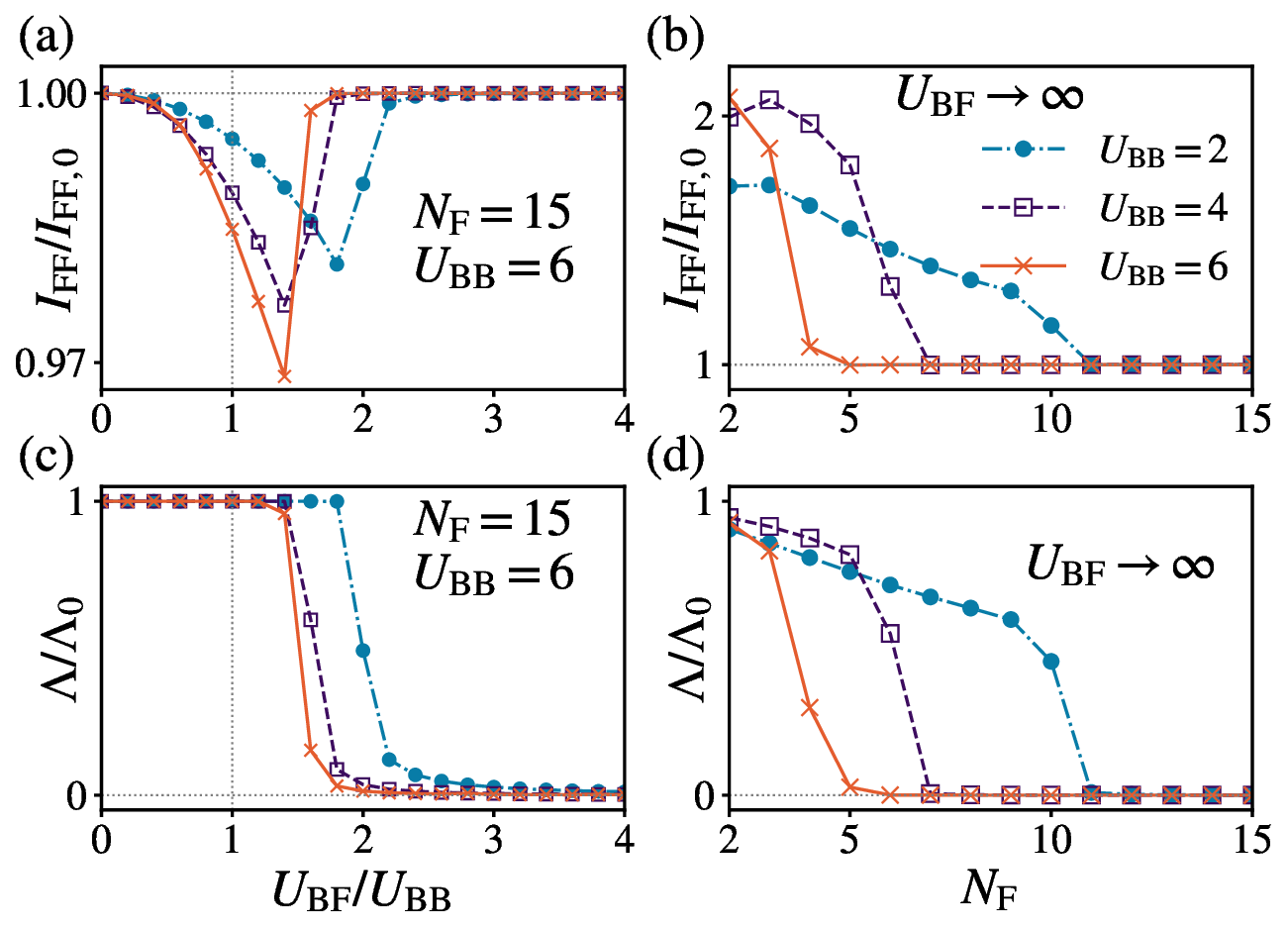}
    \caption{ 
    (a) $I_{\rm FF}$ vs $U_{\rm BF}/U_{\rm BB}$ for $N_{\rm F} = 15$ and (b) $I_{\rm FF}$ vs $N_{\rm F}$ for $U_{\rm BF} \to \infty$, with $U_{\rm BB} = 2$ (blue dot-dashed), $4$ (purple dashed) and $6$ (orange solid), and Periodic(B)-Open(F) BCs. (c) $\Lambda$ vs $U_{\rm BF}/U_{\rm BB}$ for $N_{\rm F} = 15$ and (d) $\Lambda$ vs $N_{\rm F}$ with the corresponding interaction parameters and BCs. We normalize $I_{\rm FF}$ and $\Lambda$ with their non-interacting value $I_{\rm{FF},0}=I_{\rm FF}(U_{\rm BF}=0)$, $\Lambda_{0}=\Lambda(U_{\rm BF}=0)$ and draw gray dotted lines to indicate $U_{\rm BF}= U_{\rm BB}$ (vertical), $I_{\rm FF} = I_{\rm{FF}, 0}$ and $\Lambda = \Lambda_{0}$ (horizontal) as a reference.
    } 
    \label{fig:CorrelationsImbalance}
\end{figure}

To study the correlations around the transition to the phase-separated state, the density overlap $\Lambda$ and the mutual information $I_{\rm FF}$ represent the ideal tools in our framework, as the former is zero at phase separation and the latter signals the formation of the $N$-polaron state through the presence of bath-induced correlations when $I_{\rm FF}>I_{\rm FF,0}$, with $I_{\rm FF,0}$ being the mutual information of the $N$ fermions decoupled from the bosonic background ($U_{\rm BF}=0$). We do not use $S_{\rm BF}$ since its calculation is, practically speaking, only possible for very few impurities since the dimension of $\hat \rho_{\rm F}$ increases as $L^{N_{\rm F}} \times L^{N_{\rm F}}$. To investigate the miscible-immisible transition for a large number of impurities, $N_{\rm F}=15$, we show in Fig.~\ref{fig:CorrelationsImbalance}(a) how $I_{\rm FF}/I_{\rm{FF},0}$ changes as a function of interspecies coupling $U_{\rm BF}/U_{\rm BB}$, for fixed boson-boson interactions $U_{\rm BB}=2$, $4$ and $6$.
In all cases, we observe that the transition to phase separation is similar to the Open(B)-Open(F) BCs case (Fig.~\ref{fig:CorrelationsOpenOpen}(b)), however, while the transition still occurs at larger values of the boson-fermion coupling $U_{\rm BF}>U_{\rm BB}$, the transition point tends to $U_{\rm BF}\rightarrow U_{\rm BB}$ the deeper the bosons are in the MI regime. This is supported by the overlap calculations shown in Fig.~\ref{fig:CorrelationsImbalance}(c) where $\Lambda/\Lambda_0$ decreases from $1$ (miscible) to $0$ (immiscible) around the transition point for each $U_{\rm BF}$.

Finally, we investigate the crossover between the $N$-polaron state and the phase-separated state as a function of the number of impurities $N_{\rm F}$. In Fig.~\ref{fig:CorrelationsImbalance}(b) we show $I_{\rm FF}/I_{\rm{FF},0}$ for different fixed $U_{\rm BB}$ values and in the $U_{\rm BF} \to \infty$ limit whereby we have removed double interspecies occupancies. This figure shows how a larger number of impurities inhibits the bath-mediated interactions which reduce $I_{\rm FF}$ until a critical impurity number $N_{\rm F}^{\rm c}$ is reached at $I_{\rm FF}=I_{\rm{FF},0}$, indicating that the ground state is now phase-separated. This can also be observed in Fig.~\ref{fig:CorrelationsImbalance}(d) as the overlap goes to zero at the same critical impurity number. 
We note the important role of the bosonic interactions $U_{\rm BB}$ on the critical impurity number. As can be seen from Fig.~\ref{fig:CorrelationsImbalance}(b,d), for larger boson-boson interactions, phase separation occurs even for a small number of impurities as boson-boson correlations prevent overlap with the fermionic cloud, analogous to the behavior of $S_{\rm BF}$ at large $U_{\rm BF}$ in Fig.~\ref{fig:CorrelationsOpenOpen}(a), therefore hastening the formation of the phase separated state.

\section{System size scaling} \label{sec:scaling}

So far we have shown results for a fixed number of lattice sites ($L=30$), here we explain how these extend to different system sizes while fixing the unit bosonic filling $N_{\rm B}=L$. We focus on the case $U_{\rm BB}=2$ in the limit $U_{\rm BF} \to \infty$ which will signal if the polaron state persists or not for arbitrary large interspecies interactions. 
\begin{figure}[tb]
    \centering
    \includegraphics[width=\linewidth]{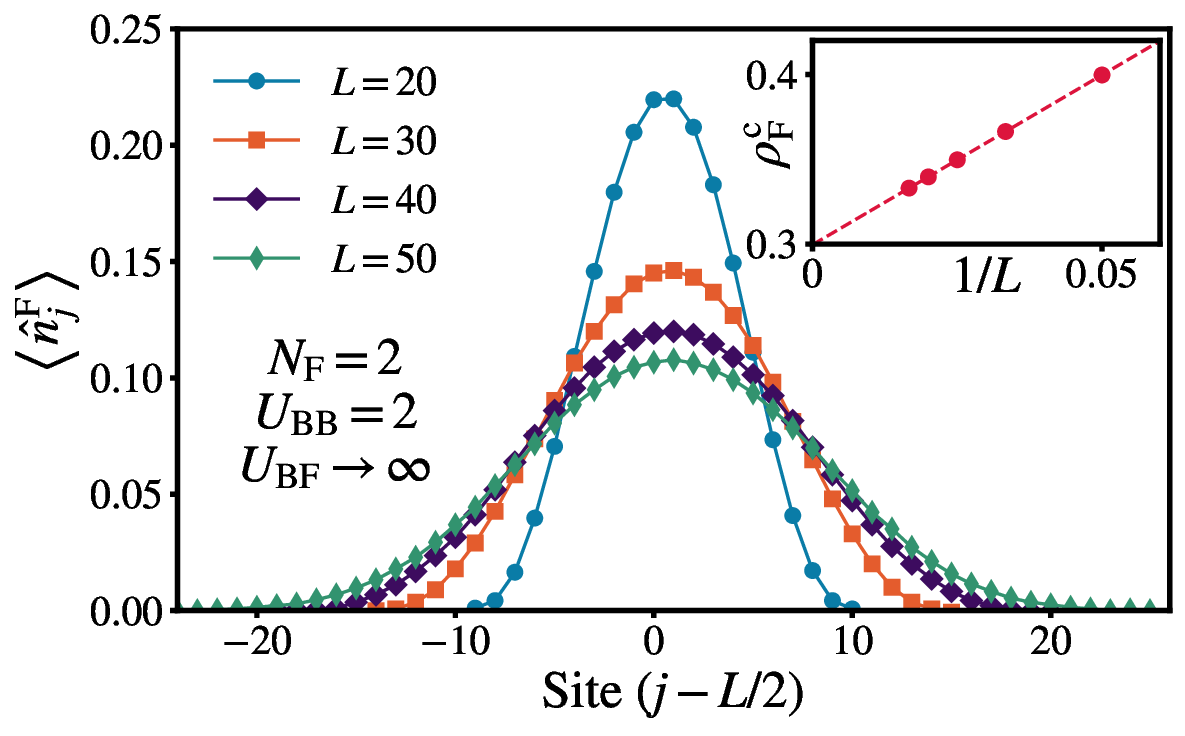}
    \caption{ Comparison of fermionic density profiles for $N_{\rm F}=2$ at different system sizes $L=N_{\rm B}$ for $U_{\rm BB}=2$ and in the $U_{\rm BF} \to \infty$ limit. Inset: Critical fermionic density $\rho_{\rm F}^{\rm c}$ for the $N$-polaron to phase separation transition as a function of $1/L$ and a linear extrapolation to the thermodynamic limit ($L \to \infty$).
    } 
    \label{fig:Scaling}
\end{figure}
In Fig.~\ref{fig:Scaling} we show the fermionic density profile for $N_{\rm F}=2$ and different system sizes $L$.
%, all of which are aligned at the center of each system. 
In all cases, we observe the characteristic density of the delocalized bipolaron state whose width scales with the increasing box size $L$ and 
%which decreases in magnitude as we increase the box size $L$ to conserve the number of impurities, showing the same physics for $N_{\rm F}=2$ as the case studied through
shows analogous features as discussed in Sec.~\ref{sec:formation}-\ref{sec:strong_regime}.

This raises the question how the polaron state will persist for increasing impurity density and system size, i.e. when does the polaron state transition to the phase separated state as shown in Fig.~\ref{fig:PopulationImbalance}. To answer this we take a similar approach as in Fig.~\ref{fig:CorrelationsImbalance}(b) where we track the critical density of impurities $\rho_{\rm F}^{\rm c}=N_F^{\rm c}/L$ after which $I_{\rm FF} = I_{\rm{FF},0}$, which indicates the phase separation transition for increasing system size $N_{\rm B}=L$. We find this to be a better criterion than the vanishing overlap since there are small contributions to $\Lambda$ in the phase separation regime for finite system sizes which complicate the characterization of the transition. In the inset of Fig.~\ref{fig:Scaling} we show the critical fermionic density $\rho_{\rm F}^{\rm c}$ as a function of the inverse system size $1/L$ to study the behavior in the thermodynamic limit ($L \to \infty$). The data shows a linear tendency which converges to a critical density $\rho_{\rm F}^{\rm c}(L\to \infty) = 0.3$, hence we conclude that for $U_{\rm BB}=2$ and $\rho_{\rm F} < \rho_{\rm F}^{\rm c}(L\to \infty)$ the $N$-polaron state can emerge for arbitrary system sizes. 
%We expect this critical density to decrease with larger bosonic intraspecies interactions as observed in Fig.~\ref{fig:CorrelationsImbalance}(b,d), which could lead to regimes where the $N$-polaron state is screened completely by the intrabath correlations, this would be an interesting future study. 

\section{Conclusions} \label{sec:conclusions}

We have studied the formation of a $N$-polaron state, where fermionic impurities are tightly bound together by a unit-filled bosonic gas in a one-dimensional optical lattice.
We have shown that a superfluid-insulator transition is crucial for polaron formation, as when the Mott phase is broken and the background becomes a superfluid, correlations with the bosonic bath bind fermions together through strong mediated attractive interactions. This allows for the redistribution of the polaron across the superfluid bath and prevents phase separation, similar to Ref.~\cite{Zschetzsche_PRR2024} and recently observed in DMRG simulations~\cite{Breu_PRA2025}. 

This also suggests that the presence of strong impurity-bath correlations could inhibit the formation of the \textit{self-pinned state}~\cite{Keller_PRL2022}, which relies on a mean-field interspecies coupling for the particles to self-localize and reorganize into a periodic structure. While the separable ansatz used at the end of Sec.~\ref{sec:strong_regime} could realize an analog of the \textit{self-pinned state} in our lattice model, using DMRG for the full Hamiltonian resulted in different physics since the bath-mediated correlations favor the delocalized polaronic state. However, in our results, we do find promising evidence that the \textit{self-pinned state} begins to emerge when the impurities self-localize before the MI-SF transition, as explained in Sec.~\ref{sec:SF_MI}. We then expect that the \textit{self-pinned state} could emerge by damping mediated correlations in the bath which can be achieved by exploring regimes of larger bosonic density, similar to the phase separation observed for large population imbalance discussed in Sec.~\ref{sec:populationimbalance}, or by tuning the mass ratio in the mixture.

%we observed similar results to our previous work that predicted this state for continuous systems, given that in both studies we only couple the mixture through the density-density interaction. Nevertheless, using DMRG for the full Hamiltonian we found different physics since bath-mediated correlations favor the delocalized polaronic state. The closest state we obtained was the onset of self-localization observed with a MI background as explained in Sec.~\ref{sec:SF_MI}, then we expect that the \textit{self-pinned state} could emerge by damping mediated correlations in the bath which can be achieved exploring regimes of larger bosonic density similar to the phase separation observed for large population imbalance discussed in Sec.~\ref{sec:populationimbalance} or tuning the mass ratio in the mixture.

We have also shown that population imbalance is key for $N$-polaron formation, and that suitable tuning of bath intraspecies interactions and correlations could allow the creation of large and stable strongly-entangled $N$-polaron states. Our analysis allowed us to determine a fermionic density threshold below which the polaron emerges that could be of use in possible experimental realizations. Indeed, a recent similar work has explored how the filling fraction and boson-boson interactions can affect bipolaron formation elucidating complementary results to our own~\cite{Dominguez-Castro_2024}. Future studies could also investigate the role of particle statistics inside the polaronic state, analog to previous work with bosonic impurities using exact diagonalization~\cite{Yordanov_JPB2023, Isaule_SciPC2024}. Moreover, the use of mixed BCs on other well-known impurity systems could lead to novel phenomena by blocking or enhancing boundary effects.

\acknowledgements
This work was supported by the Okinawa Institute of Science and Technology Graduate University. The authors are grateful for the Scientific Computing and Data Analysis (SCDA) section of the Research Support Division at OIST. The authors thank Tim Keller, L. A. Pe\~na Ardila, G. A. Dom\'inguez-Castro and F. Isaule for insightful discussions. T.F. acknowledges support from JSPS KAKENHI Grant No. JP23K03290. T.F. and T.B. are also supported by JST Grant No. JPMJPF2221.

\appendix

\section{Ground-state search using DMRG} \label{sec:DMRG}

To find the ground state of the Hamiltonian~\eqref{eq:Hubbard}-\eqref{eq:Hubbard_bf}, the Density Matrix Renormalization Group (DMRG) technique~\cite{White_PRL1992} is used in the tensor network formalism~\cite{Schollwock_AP2011} with the help of the TeNPy library~\cite{Hauschild_SciPPLN2018}. As it is necessary to truncate the local bosonic Hilbert space, the soft-core approximation is used for the maximum number of bosons per site $N_{\rm B}^{\rm{max}}=5$, which is large enough to capture the essential physics for $U_{\rm BB} \geq 2$~\cite{Pai_PRL1996, Ejima_EPL2011, Avella_PRA2020, Rossini_NJP2012} and results in a local basis of size $12$ with all of the possible combinations of fermionic and bosonic occupations under the assumed approximation. In the $U_{\rm BF}\to \infty$ approximation the local basis is reduced to $7$ elements by removing the states with a boson and a fermion on the same site.

\par%

Given that a periodic boundary condition adds a long-range hopping between the edges of the one-dimensional lattice chain, for the construction of the matrix product state (MPS) we reorder the system as $\lbrace s_1, s_L, s_2, s_{L-1}, ... \rbrace$, where $s_j$ represents the $j$-th site in the original ordering.
This increases the nearest-neighbor hopping to the next-to-nearest neighbor but reduces the long-range hopping at the edges to a nearest-neighbor term. The advantage of this transformation is that it preserves locality in the open one-dimensional system represented by the MPS which upholds the validity of the area law scaling of a subsystem's entanglement entropy with the size of its boundary, crucial for the efficiency of MPS algorithms~\cite{Eisert_RMP2010}. On the other hand, alternatives such as a pure periodic MPS~\cite{Schollwock_AP2011} where an additional bond closes the MPS by its edges require a more elaborate framework~\cite{Porras_PRB2006, Pippan_PRB2010, Weyrauch_UJP2018} since it is not possible to split the system in two with a singular value decomposition, a necessary operation for most of the MPS algorithms, hence we did not consider this alternative method to be able to use the available software for standard MPS~\cite{Hauschild_SciPPLN2018}.

\par%

During the DMRG algorithm, a maximum truncation error of the many-body Hilbert space of $10^{-14}$ is set; based on this criteria, the library optimizes the bond dimension $\chi$ along the chain with a fixed upper bound $\chi_{\rm{max}}$. To improve the algorithm's convergence, we ramp up $\chi_{\rm{max}}$ by 100 for every 10 sweeps until it reaches a maximum cap of 3000.
The program optimizes the wave function for each simulation until the ground state energy has an error less than $10^{-7}$ and the entanglement entropy at the middle of the chain less than $10^{-3}$.

\par%

We also consider the case of a separable wave function between species. To tackle this problem, we split the bosonic \eqref{eq:Hubbard_b} and fermionic \eqref{eq:Hubbard_f} Hamiltonians and minimize the energy iteratively, one component at a time, while taking into account the density-density coupling \eqref{eq:Hubbard_bf}. We use an MPS ansatz for the bosonic species to capture the correlations of the Bose Hubbard model and apply a DMRG ground-state search; for each sweep, we fix the impurity density, while between sweeps, the bosonic background is fixed, and the fermionic Hamiltonian is solved exactly, given that it is quadratic. Since the local basis for the Bose Hubbard model is smaller than the mixed case ($N_{\rm B}^{\rm{max}}+1=6$ states), we only require bond dimensions of $\chi_{\rm max} \approx 300$. The rest of the parameters are kept the same as in the complete model.

\section{Bosonic correlations in the SF-MI transition} \label{sec:SF_MI_corr}

Self-localization in this system emerges because of the difference in the spread of correlations between MI and SF backgrounds. This appendix intends to characterize this for the Periodic(B)-Open(F) BCs case as the focus of this work, by using the bosonic correlation function $\widetilde{\rho}^{(1)}_{\rm B}$ where we average the corresponding reduced density matrix at a given distance:
\begin{equation}
    \widetilde{\rho}^{(1)}_{\rm B}(d) = \frac{\sum_{j,k}^{L} \delta_{ \min\left( |j-k|, L-|j-k|\right), d }  \expval{\hat b^\dagger_k \hat b_j} }{\sum_{j,k}^{L} \delta_{ \min\left( |j-k|, L-|j-k|\right), d } }. 
\end{equation}
Here, the Kronecker delta signals the pair of sites at a distance $d$ from each other, taking into account the periodic boundary conditions of the bosonic system.

\begin{figure}[h]
    \centering
    \includegraphics[width=0.9\linewidth]{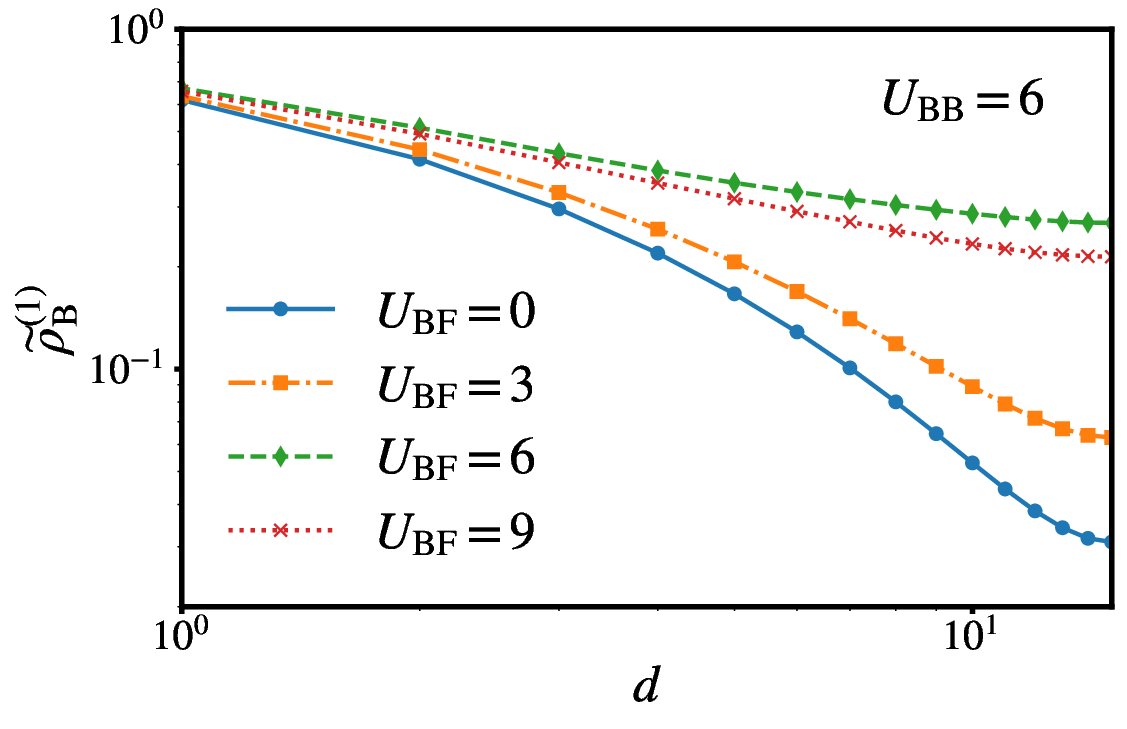}
    \caption{Averaged correlation function $\widetilde{\rho}^{(1)}_{\rm B}$ vs distance $d$ for $U_{\rm BB}=6$ and $U_{\rm BF}=0$, $3$, $6$ and $9$ with Periodic(B)-Open(F) BCs.}
    \label{fig:CorrelationDistance}
\end{figure}

We plot $\widetilde{\rho}^{(1)}_{\rm F}(d)$ in Fig.~\ref{fig:CorrelationDistance} for $U_{\rm BB}=6$ and different $U_{\rm BF}$ values. The maximum of the relative distance between the fermions signals the MI-SF transition ($U_{\rm BF} \approx 4$, see Fig.~\ref{fig:DensityMiscibilityGap}(f)), so we consider representative values of interspecies interaction below, $U_{\rm BF}=0$ and $U_{\rm BF}=3$, and above, $U_{\rm BF}=6$ and $U_{\rm BF}=9$, this transition. For $U_{\rm BF}=0$ and $U_{\rm BF}=3$ the decay of the correlation function is exponential, except for finite size effects at large $d$, a behavior characteristic of an insulating state which in turn suppresses correlation between the fermions.
Then, for interactions $U_{\rm BF}=6$ and $U_{\rm BF}=9$ larger than the transition critical value, long-range correlations in the form of an algebraic decay indicate the SF background transition where boson-mediated fermionic entanglement can be generated.

\section{Quasiparticle residue}\label{sec:quasiparticle}
\begin{figure}[tb]
    \centering
    \includegraphics[width=0.9\linewidth]{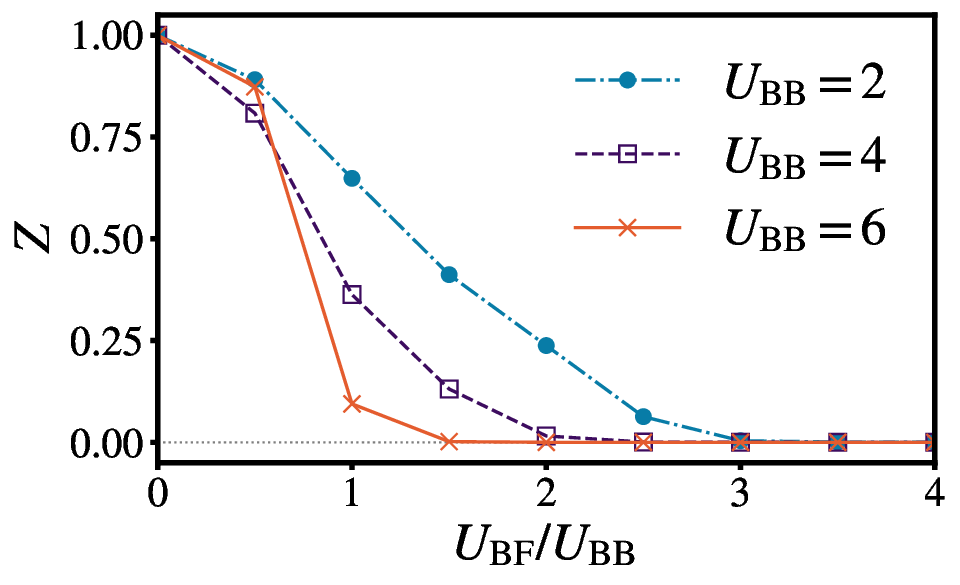}
    \caption{ Quasiparticle weight $Z$ as a function of the interspecies interaction $U_{\rm BF}$ scaled by a fixed $U_{\rm BB} = 2$, $4$ and $6$. $Z=0$ is shown as a dotted line for reference. 
    } 
    \label{fig:QuasiparticleWeight}
\end{figure}

A key concept in the analysis of polaron physics is the quasiparticle weight~\cite{Grusdt_2024}, defined as the squared overlap between the coupled ($U_{\rm BF}=0$) and uncoupled ($U_{\rm BF}\neq 0$) states. We define the quasiparticle weight through the overlap of the full wavefunctions 
    %[Grusdt \textit{et al}, arxiv.2410.09413 (2024)]
\begin{align}
    Z=|\braket{U_{\rm BF}=0, U_{\rm BB}}{U_{\rm BF},U_{\rm BB}}|^2\,,
\end{align}
with the interacting wave function $\ket{U_{\rm BF}, U_{\rm BB}}$ and the initial decoupled state $\ket{U_{\rm BF}=0, U_{\rm BB}}$. In Fig.~\ref{fig:QuasiparticleWeight}, we show this quantity as a function of the interspecies interaction for multiple $U_{\rm BB}$ values, all of which show the convergence of $Z\to 0$ as we increase $U_{\rm BF}$. We would like to stress that this quantity is strongly affected by the larger bosonic component, which dominates the overlap as $N_{\rm B}\gg N_{\rm F}$. In the regime of strong impurity-bath interaction $U_{\rm BF}/U_{\rm BB}\gg1$, the impurities strongly perturb the bosonic bath and drive it to orthogonality. This is also accelerated around a quantum critical point in the system, which we have when the impurities induce the MI-SF transition and the fermions collapse into the bound polaronic state. Crucially, the overlap will decay with increasing number of bosons, $Z\sim N_{\rm B}^{-\alpha}$, as described by the orthogonality catastrophe \cite{Anderson_PRL1967}. The vanishing of the quasiparticle residue has also been explored previously for Bose-polarons \cite{Grusdt_NJP2017} and is a characteristic of 1D systems, so we see that the quasi-particle residue vanishing in our case is to be expected.

\bibliographystyle{apsrev4-1}
\bibliography{BoseFermiImpurities} 

\end{document}